\documentclass[aip,amsmath,amssymb,reprint]{revtex4-1}

\usepackage{graphicx}
\usepackage{bm}

\usepackage{inputenc}
\usepackage[T1]{fontenc}
\usepackage{mathptmx}
\usepackage{etoolbox}
\usepackage{xcolor}
\usepackage{tabularx}
\usepackage{enumitem}
\usepackage{subcaption} 
\usepackage{hyperref}
\usepackage{calc}

\definecolor{incol}{rgb}{1,0,0}

\newcommand{\angstrom}{\text{\normalfont\AA}}
\newcommand{\cfg}{{\rm cfg}}

\makeatletter
\patchcmd{\maketitle}{\@fnsymbol}{\@alph}{}{}  
\makeatother

\begin{document}

\title[MLIP-3: A Software Package for Active Learning on Atomic Neighborhoods]{MLIP-3: Active learning on atomic environments with Moment Tensor Potentials}

\author{Evgeny Podryabinkin} 
\affiliation{Skolkovo Institute of
Science and Technology, Skolkovo Innovation Center, Bolshoy boulevard 30, Moscow, 143026, Russian Federation.}

\author{Kamil Garifullin} 
\affiliation{Moscow Institute of Physics and Technology, Russian Federation.}

\author{Alexander Shapeev} 
\affiliation{Skolkovo Institute of
Science and Technology, Skolkovo Innovation Center, Bolshoy boulevard 30, Moscow, 143026, Russian Federation.}

\author{Ivan Novikov} 
\thanks{Corresponding author: i.novikov@skoltech.ru}
\affiliation{Skolkovo Institute of
Science and Technology, Skolkovo Innovation Center, Bolshoy boulevard 30, Moscow, 143026, Russian Federation.}
\affiliation{Moscow Institute of Physics and Technology, Russian Federation.}

\date{\today}

\begin{abstract}
Nowadays, academic research relies not only on sharing with the academic community the scientific results obtained by research groups while studying certain phenomena, but also on sharing computer codes developed within the community.
In the field of atomistic modeling these were software packages for classical atomistic modeling, later---quantum-mechanical modeling, and now with the fast growth of the field of machine-learning potentials, the packages implementing such potentials.
In this paper we present the MLIP-3 package for constructing moment tensor potentials and performing their active training.
This package builds on the MLIP-2 package (Novikov et al. (2020), The MLIP package: moment tensor potentials with MPI and active learning. Machine Learning: Science and Technology, 2(2), 025002.), however with a number of improvements, including active learning on atomic neighborhoods of a possibly large atomistic simulation.
\end{abstract}

\maketitle

\section{Introduction}

The success of machine-learning potentials in recent years is revolutionizing computational materials science.
Due to their unique combination of accuracy and performance, they have enabled solving many atomistic modeling problems that were previously impossible or impractical to solve due to excessive consumption of computational resources \cite{gubaev2019-alloys,novikov2018-rpmd,mortazavi2021-mtp-shengbte,podryabinkin2022-nanohardness,novoselov2019-diffusion}.
Machine-learning potentials are based on the idea of training a classical (as opposed to quantum-mechanical) interatomic interaction model on the results of first-principles calculations and further reproducing the behavior of this model in practical simulations. The feature that distinguishes machine-learning potentials from semiempirical-based models with coefficients found from first-principles calculations is that their flexible functional form allows the potential energy surface to be approximated with arbitrary accuracy by increasing the number of parameters, and consequently increasing the computation cost, however keeping it orders of magnitude below the cost of first-principles calculations.

There is a number of machine-learning potentials developed, including SNAP---the potentials on the basis of spherical harmonics \cite{thompson2015-automated}, GAP---the potentials based on Gaussian processes \cite{bartok2010-GAP}, ANN---the potentials based on artificial neural networks \cite{behler2007-NNP}, and MTP---the polynomial-like potentials based on the tensors of inertia of atomistic environments \cite{shapeev2016-mtp}.
Examples of more recent potentials are polynomial-like ACE potentials \cite{drautz2019-ACE, dusson2022-ACE}, and neural network-based DeepMD \cite{wang2018-deepmd}, PINN \cite{pun2019-pinn}, and NeQUIP \cite{batzner2022-equivarnn}.

In \cite{zuo2020-benchmark}, an independent benchmarking of the first four potentials was carried out, in which MTPs (moment tensor potentials) showed a very favorable performance in terms of accuracy versus computational efficiency. The MTPs and tools to work with them have been implemented and made available for use in the software package MLIP-2 \cite{novikov2020-mlip-2}.
The MTP potentials has been used in solving a number of materials science problems including predicting crystal structures \cite{podryabinkin2019-uspex}, searching for new materials and high-performance screening of structures \cite{gubaev2019-alloys}, modeling chemical reactions \cite{novikov2018-rpmd,novikov2019-rpmd-SH2}, calculating melting points \cite{novikov2020-mlip-2}, thermal conductivity \cite{mortazavi2021-mtp-shengbte, mortazavi2022-bc2n}, hardness \cite{podryabinkin2022-nanohardness}, diffusion coefficients \cite{novoselov2019-diffusion}, and other computationally expensive properties of materials\cite{Shapeev_2020_elinavar}.

Three years have passed since the MLIP-2 code was published, and during that time, a number of changes and improvements has been has been implemented in the new version of the code. The changes are related to simplification of operations with machine-learning potentials, including their active learning, and new capabilities of parallel calculations. Besides that, we present a new feature to train MTPs on local fragments of a large atomistic configuration in the new software version. The purpose of this article is to introduce the reader to the MLIP-3 package and to describe its features. It should be noted that the functional form of MTPs, as well as their learning and extrapolation detection algorithms, remained essentially unchanged as compared to the MLIP-2 package \cite{novikov2020-mlip-2}. At the same time, the user interface underwent substantial revision, new parallelism capabilities were added, and perhaps most importantly, the possibility of active learning on atomistic environments (neighborhoods) was implemented.

\section{Theory}

\subsection{Moment Tensor Potential}

Moment Tensor Potential (MTP) is a machine-learning interatomic potential which was first proposed in \cite{shapeev2016-mtp} for single-component materials and its generalizations were described in \cite{gubaev2018-chemoinformatics, novikov2022-magneticMTP}. MTP is a local potential, i.e., the energy $E^{\rm MTP}$ is the sum of contributions $V^{\rm MTP}(\mathfrak{\bm n}_i)$ of atomic neighborhoods ${\bf \mathfrak{n}}_i$ for $N$ atoms
\begin{align} \label{EnergyMTP}
E^{\rm MTP} = \sum \limits_{i=1}^{N} V^{\rm MTP}(\mathfrak{\bm n}_i). 
\end{align}
Each neighborhood is a tuple $\mathfrak{ n}_i = ( \{r_{i1},z_i,z_1 \}, \ldots, \{r_{ij},z_i,z_j \}, \ldots, \{r_{iN_ {\rm nbh}},z_i,z_{N_ {\rm nbh}} \} )$, where $r_{ij}$ are relative atomic positions, $z_i$, $z_j$ are the types of central and neighboring atoms, and $N_ {\rm nbh}$ is the number of atoms in neighborhood. We also denote the maximum number of atomic types occurred in all the neighborhoods by $N_{\rm types}$. Each contribution $V^{\rm MTP}(\mathfrak{\bm n}_i)$ in the potential energy $E^{\rm MTP}$ expands through a set of basis functions
\begin{align} \label{SiteEnergyMTP}
V^{\rm MTP}({\bf \mathfrak{n}}_i) = \sum \limits_{\alpha=1}^{N_{\rm lin}} \xi_{\alpha} B_{\alpha}({\mathfrak{\bm n}}_i),
\end{align} 
where $B_{\alpha}$ are the MTP basis functions, $\xi_{\alpha}$ are the linear parameters to be found, and $N_{\rm lin}$ is the number of these parameters. To define the functional form of the MTP basis functions and the number $N_{\rm lin}$ we introduce the so-called moment tensor descriptors
\begin{equation}\label{MomentTesnsorDescriptors}
M_{\mu,\nu}({\mathfrak{\bm n}}_i)=\sum_{j=1}^{N_{\rm nbh}} f_{\mu}(|r_{ij}|,z_i,z_j) r_{ij}^{\otimes \nu}.
\end{equation}
The descriptor consists of the angular part $r_{ij}^{\otimes \nu}$ (the symbol ``$\otimes$'' denotes the outer product of vectors and, thus, the angular part is the tensor of $\nu$-th order) and the radial part $f_{\mu}(|r_{ij}|,z_i,z_j)$ of the following form
\begin{align} \label{RadialFunction}
\displaystyle
f_{\mu}(|r_{ij}|,z_i,z_j) = \sum_{\beta=1}^{N_{\rm polyn}} c^{(\beta)}_{\mu, z_i, z_j} T^{(\beta)} (|r_{ij}|) (R_{\rm cut} - |r_{ij}|)^2.
\end{align}
Here $\mu=0, \ldots, N_{\rm rad}-1$ is the number of the radial function $f_{\mu}$ (the method to define a concrete number $N_{\rm rad}$ of radial functions needed to construct all the MTP basis functions is detailed below), $c^{(\beta)}_{\mu, z_i, z_j}$ are the radial parameters to be found, $T^{(\beta)} (|r_{ij}|)$ are polynomial functions, $N_{\rm polyn}$ is their number. The number of the radial MTP parameters $c^{(\beta)}_{\mu, z_i, z_j}$ is $N_{\rm polyn} \times N_{\rm rad} \times N_{\rm types}^2$. Finally, the term $(R_{\rm cut} - |r_{ij}|)^2$ is introduced to ensure smoothness with respect to the atoms leaving and entering the sphere with the cut-off radius $R_{\rm cut}$.
 
By definition, the MTP basis function $B_{\alpha}$ is a contraction of one or more moment tensor descriptors, yielding a scalar, e.g.
$$M_{1,0}, ~(M_{1,2} M_{0,1}) \cdot M_{0,1}, ~M_{1,2} : M_{3,2}, \ldots,$$
where ``$\cdot$'' is an inner product of vectors, ``$:$'' is a Frobenius product of matrices. In order to construct the basis functions $B_{\alpha}$, and, thus, determine a particular functional form of MTP, we define the so-called \emph{level} of moment tensor descriptor
\begin{equation} \label{eq:LevelMTD}
\displaystyle
{\rm lev} M_{\mu,\nu} = 2 + 4 \mu + \nu,
\end{equation}
for example, ${\rm lev} M_{0,0} = 2$, ${\rm lev} M_{2,1} = 11$, ${\rm lev} M_{0,3} = 5$. We also define the level of MTP basis function 
\begin{equation} \label{LevelMultMTD}
\displaystyle
{\rm lev} B_{\alpha} = \rm {lev} \underbrace {\prod_{p=1}^{P} M_{\mu_p,\nu_p}}_{scalar} = \sum \limits_{p=1}^P (2 + 4 \mu_p + \nu_p).
\end{equation}
A set of MTP basis functions and, thus, a particular functional form of MTP depends on the maximum level, ${\rm lev_{\rm max}}$, which we also call the level of MTP. We include in the set of the MTP basis functions only the ones with ${\rm lev} B_{\alpha} \leq {\rm lev_{\rm max}}$, e.g., the MTP of 6-th level includes five basis functions:
$$B_1=M_{0,0}; ~{\rm lev} {B_1} = 2 \le \rm {lev}_{\rm{max}} = 6,$$
$$B_2=M_{1,0}; ~{\rm lev} {B_2} = 6 \le \rm {lev}_{\rm{max}} = 6,$$ 
$$B_3=M_{0,0}^2; ~{\rm lev} {B_3} = 4 \le \rm {lev}_{\rm{max}} = 6,$$ 
$$B_4=M_{0,1} \cdot M_{0,1}; ~{\rm lev} {B_4} = 6 \le \rm {lev}_{\rm{max}} = 6,$$  
$$B_5=M_{0,0}^3; ~{\rm lev} {B_5} = 6 \le \rm {lev}_{\rm{max}} = 6.$$
Thus, the number of linear parameters $N_{\rm lin}$ and the number $N_{\rm rad}$ of radial functions depend on the level of MTP and are fixed for each level (e.g., if ${\rm {lev}_{\rm{max}}} = 6$ then $N_{\rm lin} = 5$, $N_{\rm rad} = 2$). 

\subsection{Fitting}

For finding MTP parameters we fit it on a (quantum-mechanical) training set. Let $K$ be a number of configurations in the training set, and $N^{(k)}$ is a number of atoms in $k$-th configuration. Denote a vector of MTP parameters to be found by ${\bm \theta} = ( \xi_{\alpha}, c^{(\beta)}_{\mu, z_i, z_j})$. We find the optimal parameters ${\bm \bar{\theta}}$ by solving the following optimization problem (minimization of the objective function)
\begin{equation} \label{Fitting}
\begin{array}{c}
\displaystyle
\sum \limits_{k=1}^K \Bigl[ w_{\rm e} \left(E^{\rm MTP}_k({\bm {\theta}}) - E^{\rm QM}_k \right)^2 + w_{\rm f} \sum_{i=1}^{N^{(k)}} \left| {\bf f}^{\rm MTP}_{i,k}({\bm {\theta}}) - {\bf f}^{\rm QM}_{i,k} \right|^2 
\\
\displaystyle
+ w_{\rm s} \sum_{a,b=1}^3 \left( \sigma^{\rm MTP}_{ab,k}({\bm {\theta}}) - \sigma^{\rm QM}_{ab,k} \right)^2 \Bigr] \to \operatorname{min}.
\end{array}
\end{equation} 
We start from randomly initialized MTP parameters. The optimal parameters ${\bm \bar{\theta}}$ are found numerically, by using the iterative method for minimization of the nonlinear objective function, namely, Broyden-Fletcher-Goldfarb-Shanno algorithm \cite{Fletcher1991_BFGS}. Thus, after the optimization, the parameters ${\bm \bar{\theta}}$ are near the local minimum of the objective function. In this objective function, $E^{\rm QM}_k$, ${\bf f}^{\rm QM}_{i,k}$, and $\sigma^{\rm QM}_{ab,k}$ are the reference energies, forces, and stresses, i.e. the ones to which we fit the MTP energies $E^{\rm MTP}_k$, forces ${\bf f}^{\rm MTP}_{i,k}$, and stresses $\sigma^{\rm MTP}_{ab,k}$, and, thus, optimize the MTP parameters ${\bm \theta}$. The factors $w_{\rm e}$, $w_{\rm f}$, and $w_{\rm s}$ are non-negative weights which express the importance of energies, forces, and stresses with respect to each other. We refer to the minimization problem \eqref{Fitting} as the fitting of MTP.

\subsection{Active Learning: query strategy on atomic configurations}

We introduce a concept of \emph{extrapolation grade} on the basis of the papers \cite{podryabinkin2017-AL,gubaev2018-chemoinformatics}.
To that end, we start from \eqref{Fitting}.
Assume that we know the vector of optimal MTP parameters ${\bm \bar{\theta}}$ and let the length of the vector be $m$. Then we can linearize each term in the objective function, in particular, the term with the difference between energies
$$E^{\rm QM}_k - E^{\rm MTP}_k({\bm {\theta}}) \approx E^{\rm QM}_k - \sum \limits_{p=1}^{m} (\theta_p - \bar{\theta}_p) \dfrac{\partial E^{\rm MTP}_k({\bm {\bar{\theta}}})}{\partial \theta_p}.$$
We can consider the fitting of MTP as the solution of the overdertemined system
\begin{align} \label{OverdeterminedSystem1}
\sum \limits_{p=1}^{m} \theta_p \dfrac{\partial E^{\rm MTP}_k({\bm {\bar{\theta}}})}{\partial \theta_p} = E^{\rm QM}_k + \sum \limits_{p=1}^{m} \bar{\theta}_p \dfrac{\partial E^{\rm MTP}_k({\bm {\bar{\theta}}})}{\partial \theta_p}.
\end{align}
The matrix of the system \eqref{OverdeterminedSystem1} is 
\[
\mathsf{B}=\left(\begin{matrix}
\frac{\partial E^{\rm MTP}_1}{\partial \theta_1}({\bm {\bar{\theta}}}) & \ldots & \frac{\partial E^{\rm MTP}_1}{\partial \theta_m}({\bm {\bar{\theta}}}) \\
\vdots & & \vdots \\
\frac{\partial E^{\rm MTP}_K}{\partial \theta_1}({\bm {\bar{\theta}}}) & \ldots & \frac{\partial E^{\rm MTP}_K}{\partial \theta_m}({\bm {\bar{\theta}}}) 
\end{matrix}\right).
\]
From this set of equations, we select a set of the $m$ most linearly independent equations, that is, the equations with the maximum absolute value of the determinant.
This is the core idea of our active learning strategy, which is based on the D-optimality criterion. According to this criterion, the configurations that give the most linearly independent equations (maximizing absolute value of determinant) are selected and included to the training set. We call the selected set of $m$ configurations the {\it active set}, which the set of the most ``extreme'' and diverse configurations from the point of view of MTP that is being fitted. 

To construct an active set from a set of configurations, we use the MaxVol algorithm \cite{goreinov2010-maxvol}, providing a computationally efficient way to calculate how many times the absolute value of determinant of a square matrix $\mathsf{A}$ associated with the active set will change, if some configuration from the active set is replaced by some another. Let 
\[
\mathsf{b}=\left(\frac{\partial E^{\rm MTP}}{\partial \theta_1}({\bm {\bar{\theta}}}) \ldots \frac{\partial E^{\rm MTP}}{\partial \theta_m}({\bm {\bar{\theta}}}) \\
\right)
\]
the coefficients in the equation \eqref{OverdeterminedSystem1} calculated for a candidate configuration $\rm cfg$. Then the largest increase of $|\rm det(\mathsf{A})|$ is 
\begin{align*}
\gamma(\rm {cfg}) &=\max_{1\leq j \leq m} |c_j|,\qquad\text{where}
\\
\begin{pmatrix}
c_1 & \ldots & c_m
\end{pmatrix} &=
\left(\frac{\partial E^{\rm MTP}}{\partial \theta_1}({\bm {\bar{\theta}}}) \ldots \frac{\partial E^{\rm MTP}}{\partial \theta_m}({\bm {\bar{\theta}}}) 
\right) \mathsf A^{-1}.
\end{align*}
Thus, if $\gamma(\rm cfg)>1$ then $|\rm det(A)|$ could be increased $\gamma$ times when a configuration from the active set associated with index $j_0={\rm argmax_j}(c_j)$ is replaced by $\rm cfg$.

Note that if the active set and the training set are the same, then the MTP parameters can be found as 
$$
{\bm \xi} = \left(E^{\rm qm}(\cfg_1) \, \ldots\, E^{\rm qm}(\cfg_m) \right) \mathsf{A^{-1}}
$$
and the energy of any configuration can be expressed as a linear combination of energies of active set configurations
\[
E^{\rm mtp}(\cfg) = \sum_{j=1}^m c_j  E^{\rm qm}(\cfg_j),
\]
and $\gamma(\rm cfg) = \max_j |c_j|$.
Therefore we say that MTP extrapolates if $\gamma(\rm cfg)>1$ on $\rm cfg$, and interpolates otherwise. Hence $\gamma$ can be interpreted as the \textit{extrapolation grade}. 

\subsection{Active Learning: query strategy on local atomic neighborhoods}

In the previous section we formulated the strategy of active learning on atomistic configurations. This strategy works fine if we deal with relatively small configurations, which can be calculated with quantum-mechanical methods spending reasonable computational resources.
However, this becomes impractical when we deal with relatively large configurations, because their quantum-mechanical calculations are too computationally expensive.
Moreover, MTP may, in some sense, extrapolate only on some small part of the large configuration, while the rest of the configuration can be described well by the existing training set.
In this case, it may be desirable to use the active learning strategy on local atomic neighborhoods.

Before we formulate this strategy, we note that the active learning on atomistic configurations does not use the information in right part of \eqref{OverdeterminedSystem1} (i.e., does not use the quantum-mechanical energies).
Thus, if we assume that we are able to somehow calculate individual atomic energy $V^{\rm QM}(\mathfrak{\bm n}_i^{(k)})$, then we can write the overdetermined system of equations for each atomic neighborhood in every configuration from the training set:
\begin{align} 
V^{\rm MTP}(\mathfrak{\bm n}_i^{(k)}) = V^{\rm QM}(\mathfrak{\bm n}_i^{(k)}), 1\leq i \leq N^{(k)}, 1\leq k \leq K.
\end{align}

After linearization with respect to the parameters $\theta$ near the optimal values $\bar{\theta}$ we obtain 
\begin{align} \label{OverdeterminedSystem2}
\sum \limits_{p=1}^{m} \theta_p \dfrac{\partial V^{\rm MTP}({\bm {\bar{\theta}}}, \mathfrak{\bm n}_i^{(k)})}{\partial \theta_p} =  V^{\rm QM}_{i,k} + \sum \limits_{p=1}^{m} \bar{\theta}_p \dfrac{\partial V^{\rm MTP}({\bm {\bar{\theta}}}, \mathfrak{\bm n}_i^{(k)})}{\partial \theta_p} ,  1 \leq k \leq K.
\end{align}
The matrix of this system \eqref{OverdeterminedSystem2} is
\[
\mathsf{B}=\left(\begin{matrix}
\frac{\partial V^{\rm MTP}}{\partial \theta_1}({\bm {\bar{\theta}}}, \mathfrak{\bm n}_1^{(1)}) & \ldots & \frac{\partial V^{\rm MTP}}{\partial \theta_m}({\bm {\bar{\theta}}}, \mathfrak{\bm n}_1^{(1)}) \\
\vdots & & \vdots \\
\frac{\partial V^{\rm MTP}}{\partial \theta_1}\left({\bm {\bar{\theta}}}, \mathfrak{\bm n}_{N^{(1)}}^{(1)} \right) & \ldots & \frac{\partial V^{\rm MTP}}{\partial \theta_m} \left({\bm {\bar{\theta}}}, \mathfrak{\bm n}_{N^{(1)}}^{(1)} \right) \\
\vdots & & \vdots \\
\frac{\partial V^{\rm MTP}}{\partial \theta_1}({\bm {\bar{\theta}}}, \mathfrak{\bm n}_1^{(K)}) & \ldots & \frac{\partial V^{\rm MTP}}{\partial \theta_m}({\bm {\bar{\theta}}}, \mathfrak{\bm n}_1^{(K)}) \\
\vdots & & \vdots \\
\frac{\partial V^{\rm MTP}}{\partial \theta_1}\left({\bm {\bar{\theta}}}, \mathfrak{\bm n}_{N^{(K)}}^{(K)} \right) & \ldots & \frac{\partial V^{\rm MTP}}{\partial \theta_m} \left({\bm {\bar{\theta}}}, \mathfrak{\bm n}_{N^{(K)}}^{(K)} \right)
\end{matrix}\right),
\]
where the first $N^{(1)}$ rows correspond to the first configuration (of $N^{(1)}$ atoms) in the training set, the last $N^{(K)}$ rows correspond to the last ($K$-th) configuration, and, thus, the size of the matrix is $\left( \sum \limits_{k=1}^{K} N^{(k)} \right) \times m$. 

The active learning strategy on atomic neighborhoods can be formulated in the same manner as active learning on configurations. The active set in this case consist of $m$ atomistic neighborhoods, while the matrix of maximal volume $\mathsf{A}$ (i.e., square matrix with maximal absolute value of determinant) contain the corresponding elements $\frac{\partial V^{\rm MTP}}{\partial \theta_p} \left({\bm {\bar{\theta}}}, \mathfrak{\bm n}_{i}^{(k)} \right)$, where $\mathfrak{\bm n}_{i}^{(k)}$ is the neighborhood of the $i$-th atom in the $k$-th configuration. 

Given a configuration $\cfg^*$ we checked if it has an extrapolative neighborhood in the following way. We calculate the matrix   
\[
\mathsf{C}=\left(\begin{matrix}
\frac{\partial V^{\rm MTP}}{\partial \theta_1}({\bm {\bar{\theta}}}, \mathfrak{\bm n}_1^{*}) & \ldots & \frac{\partial V^{\rm MTP}}{\partial \theta_p}({\bm {\bar{\theta}}}, \mathfrak{\bm n}_1^{*}) & \ldots & \frac{\partial V^{\rm MTP}}{\partial \theta_m}({\bm {\bar{\theta}}}, \mathfrak{\bm n}_1^{*}) \\
\vdots & & \vdots & & \vdots \\
\frac{\partial V^{\rm MTP}}{\partial \theta_1}\left({\bm {\bar{\theta}}}, \mathfrak{\bm n}_{i}^{*} \right) & \ldots & \frac{\partial V^{\rm MTP}}{\partial \theta_p}({\bm {\bar{\theta}}}, \mathfrak{\bm n}_i^{*}) & \ldots & \frac{\partial V^{\rm MTP}}{\partial \theta_m} \left({\bm {\bar{\theta}}}, \mathfrak{\bm n}_{i}^{*} \right) \\
\vdots & & \vdots & & \vdots \\
\frac{\partial V^{\rm MTP}}{\partial \theta_1}\left({\bm {\bar{\theta}}}, \mathfrak{\bm n}_{N}^{*} \right) & \ldots & \frac{\partial V^{\rm MTP}}{\partial \theta_p}({\bm {\bar{\theta}}}, \mathfrak{\bm n}_N^{*}) & \ldots & \frac{\partial V^{\rm MTP}}{\partial \theta_m} \left({\bm {\bar{\theta}}}, \mathfrak{\bm n}_{N}^{*} \right) \\
\end{matrix}\right) \mathsf{A}^{-1},
\]
The maximal by absolute value element $\gamma_i = \gamma(\mathfrak{\bm n}_i^{*}) = \max_{1 \leq p \leq m} (|C_{i,p}|)$ in each row $i$ is the extrpolation degree for the corresponding neighborhood $\mathfrak{\bm n}_{i}^{*}$. If $\gamma_i > 1$ then MTP will extrapolate on the corresponding neighborhood $\mathfrak{\bm n}_{i}^{*}$.   

\subsection{Active learning on-the-fly}

\begin{figure}[ht]
	\centering
	\includegraphics[width=250pt]{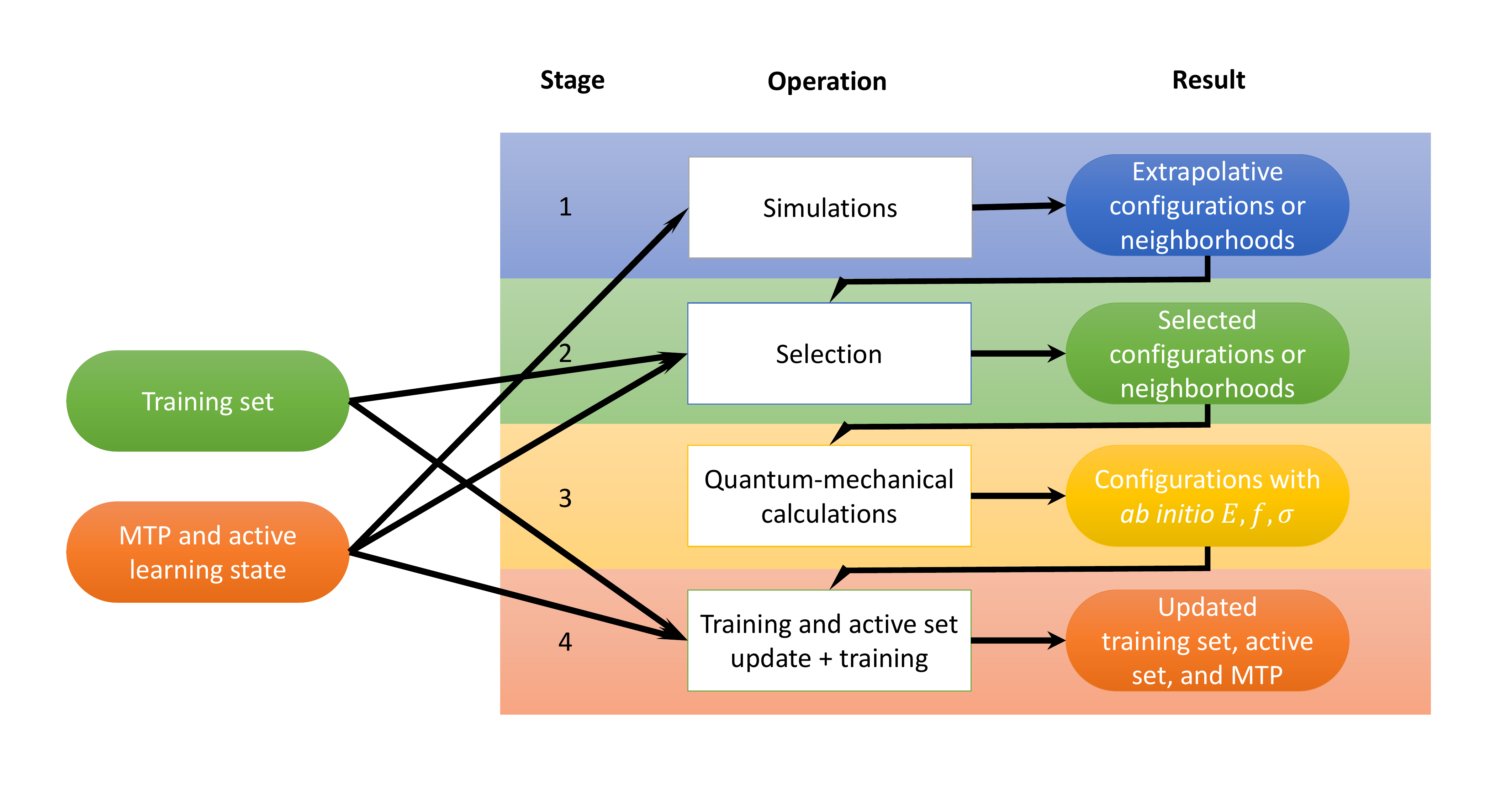}
	\caption{Scheme of active learning bootstrapping iterations.}
	\label{fig:lotf}
\end{figure}

According to our experience, the learning on-the-fly scheme described in this section is one of the most efficient method to fit an MTP for many applications.
It allows accumulating only relevant configurations in the training set which often results in optimal training errors because this MTP interpolates in the minimal required phase space. According to this scheme we start with an MTP initially trained on some (typically minimal) amount of relevant configurations and repeat the following steps (see also Fig. \ref{fig:lotf}) until the desired simulations (e.g,. MD simulations with a range of temperaturse and pressures) is completed without encountering extrapolative configurations or neighborhoods. 

\begin{enumerate}
    \item {\bf Run the simulation with extrapolation control.} The simulation algorithm generates configurations, for which energy and/or forces are computed. Prior to this calculation each configuration are assessed for extrapolation according to one of the two aforementioned active leaning strategy. That is, extrapolation grade is calculated either for the entire configuration or for each neighborhood in this configuration. In the latter case the extrapolation degree of configuration is defined as the maximal among extrapolation grades of its all neighborhoods $\gamma({\rm cfg})= \rm max_i \gamma(\mathfrak{\bm n}_i)$. Depending on the degree of extrapolation $\gamma({\rm cfg})$ we can (i) either proceed with calculation of the energy and/or forces in the case of interpolation or insignificant extrapolation, $\gamma({\rm cfg}) < \gamma_{\rm save}$; or (ii) save the configuration or its fragment to the file, and proceed with the calculation of the energy and/or forces, if extrapolation is significant but not critically large $\gamma_{\rm save} < \gamma({\rm cfg}) < \gamma_{\rm break} \approx 10$; or (iii) break the calculation if extrapolation degree exceeds the critical threshold $\gamma({\rm cfg}) > \gamma_{\rm break}$ (see Fig. \ref{fig:lotf}).

\begin{figure}[ht]
	\includegraphics[width=250pt]{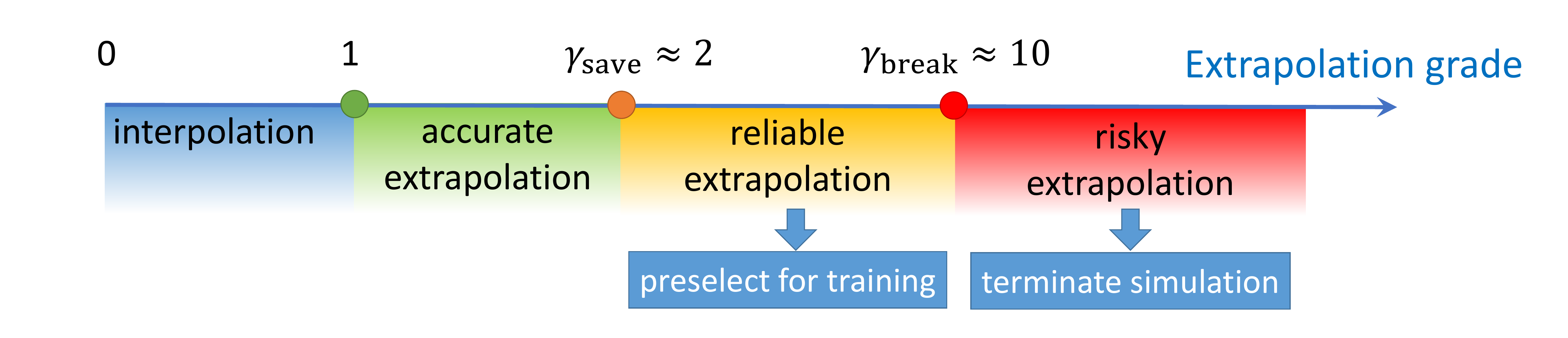}
	\caption{Classification of extrapolation grades. A configuration with the grade less than $\gamma_{\rm save}$ does not trigger any active learning actions, a configuration whose grade is between $\gamma_{\rm save}$ and $\gamma_{\rm break}$ are considered reliable yet useful for extension of the training set, but those with the grade larger than $\gamma_{\rm break}$ are risky and trigger termination of the simulation.}
	\label{Fig:thresholds}
\end{figure}

    \item {\bf Selection of configurations or neighborhoods to be learned.} The extrapolative configurations or the fragments of configurations with extrapolative environments saved to the file may be learned and improve reliability (transferability) of the potential. However the saved files may contain a large amount of similar configurations or neighborhoods. According to our active learning strategy we select among the extrapolative configurations or neighborhoods those that maximize the volume of the matrix associated with the active set.
    
    \item {\bf Quantum-mechanical calculations.} After selecting the appropriate configurations or neighborhoods for learning, it is necessary to calculate their quantum-mechanical energy, forces and stresses. In the case of active leaning on configurations, the selected configurations are calculated with the quantum-mechanical model. In the case of active leaning on neighborhoods, there are three ways processing the selected neighborhoods. (i) The simplest way is to directly calculate the configurations containing selected neighborhoods with the quantum-mechanical model. However, this option is typically impractical since these configurations are too large to be treated with the quantum-mechanical model. Hence we can either (ii) cut the extrapolative neighborhood from the configurations and treat them as non-periodic clusters with the quantum-mechanical model, or (iii) construct periodic structure from each extrapolative neighborhood in manual or semi-manual manner.
    Option (ii) is implemented in our MLIP-3 package, whereas for option (iii) the users should write their own neighborhood processing code.
    
    \item {\bf MTP re-training.} After quantum-mechanical calculations the training data calculated at the previous step, it is added to the training set, updates active set and the potential is re-trained. 
\end{enumerate}

\subsection{Protocol for choosing DFT parameters for neighborhood calculations}

In the case of active learning on atomic neighborhoods when using the DFT code with periodic boundary conditions (e.g., VASP) for calculating atomic neighborhoods the following steps for choosing DFT parameters should be done.

\begin{itemize}
    \item Choose the radius of the neighborhood (the radius should be greater than the cutoff radius of the potential) and cut the neighborhood from the large configuration.
    \item Choose 1$\times$1$\times$1 k-mesh (use this k-mesh for the neighborhood DFT calculations) and the energy cutoff greater than the energy cutoff in pseudopotentials. Add vacuum along each direction and investigate the potential energy convergence with respect to vacuum size. Find optimal vacuum size.
    \item Investigate the potential energy convergence with respect to the energy cutoff with the fixed optimal vacuum size along each direction. Find the optimal energy cutoff.
    \item Take the optimal vacuum size and the energy cutoff and investigate the convergence of the force on the central atom in the neighborhood with respect to the radius of the neighborhood. Thus choose the radius of the neighborhood. 
\end{itemize}

\section{MLIP-3 package description}

MLIP-3 is an open-source software package available at \verb|https://gitlab.com/ashapeev/mlip-3|.
It provides the code compiled to the executable file \verb|mlp| implementing a number of commands manipulating with machine-learning potentials and quantum-mechanical database.
Additionally, the plugin for using these potentials with the LAMMPS package \cite{plimpton1995-LAMMPS} is available at 
\begin{widetext}
\begin{verbatim}
https://gitlab.com/ivannovikov/interface-lammps-mlip-3/
\end{verbatim}
\end{widetext}

\subsection{MLIP-3 files and commands}

The atomic configurations are stored in the MLIP's native \verb|.cfg| files in a plain-text format. These files contain size of configuration, atomic numbers, types, positions (in $\angstrom$), supercell (in $\angstrom$), and (optionally) energies (in eV), forces (in eV/$\angstrom$), stresses (in eV), features (additional textual attributes of configuration, e.g. extrapolation grade), and nbh\_grades (or, extrapolation grade per atom, needed for query strategy on atomic neighborhoods). The reader can find more details on \verb|.cfg| format in \cite{suppinfo-manual}. 

Training set is written in \verb|.cfg| files, i.e., the configurations obtained with the codes for quantum-mechanical calculations (e.g., DFT calculations) should be converted from the formats of these codes to the native MLIP \verb|.cfg| before fitting of MTP. For example, we convert the configurations in the VASP format to the MLIP format by executing
\begin{verbatim}
mlp convert OUTCAR out.cfg --input_format=outcar    
\end{verbatim}
where \verb|OUTCAR| is the VASP output file (which is the input file for the \verb|convert| command) with energies, forces, and stresses calculated, and \verb|out.cfg| is the output file with the energies, forces, and stresses saved in the internal MLIP \verb|.cfg| format. The atomic types written in \verb|out.cfg| are relative after executing the above command, i.e., they do not correspond to the numbers of the elements in the periodic table. The \verb|absolute_types| option allows us writing absolute atomic types in the \verb|.cfg| file while converting from \verb|OUTCAR| instead of, e.g., $0,1,2,\ldots$. 

Templates of MTPs corresponding to different levels of MTP are stored in the MLIP's native \verb|.almtp| files. The MLIP-3 package contains untrained potentials defining MTPs of level $6, \ldots, 28$ in the folder \verb|MTP_templates/| of the main package folder. The templates of untrained MTPs are described in \cite{suppinfo-manual} in detail. To fit an MTP we run the following command
\begin{verbatim}
mlp train in.almtp train.cfg --save_to=pot.almtp
\end{verbatim}
The \verb|in.almtp| file contains the initial MTP, but it does not have to contain parameters in which case we start from the randomly initialized parameters. The \verb|--save_to=pot.almtp| option indicates that the trained MTP will be saved to the \verb|pot.almtp| file (if this option is skipped \verb|in.almtp| file will be overwritten). This file contains the optimal MTP parameters, the query strategy (``on configurations'' or ``on local atomic neighborhoods''), the matrix corresponding to the active set, and the configurations from the active set. We further refer the query strategy ``on configurations'' to as the ``configuration-based'' active learning mode and the query strategy ``neighborhoods-based'' to as the ``local atomic neighborhoods'' active learning mode. The default active learning mode is ``configuration-based'', but it can be specified with the option \verb|--al_mode=cfg| or \verb|--al_mode=nbh|. Active learning mode is set up at the first training and should not be changed after. 

We emphasize that after training, the \verb|.almtp| file always contains information on the active learning mode and the active set, i.e., all the data related to active learning will be automatically updated after each re-training. However, MTPs can also be used in the passive learning mode. In this case the data related to active learning will simply be ignored. 

We check the fitting errors by executing
\begin{verbatim}
mlp check_errors pot.almtp train.cfg
\end{verbatim}
The computed errors will be reported to the screen. 

Calculation of the energies, forces, and stresses for the configurations in a file \verb|in.cfg| is done by 
\begin{verbatim}
mlp calculate_efs pot.almtp in.cfg efs.cfg
\end{verbatim}

For estimation of extrapolation grades of configurations/per atoms the following command is used:
\begin{verbatim}
mlp calculate_grade pot.almtp in.cfg grade.cfg
\end{verbatim}
We note that all the data needed for these commands will be loaded from the files \verb|pot.almtp| and \verb|in.cfg|.

The trained MTP can be used for different types of atomistic simulations, e.g., relaxation (minimization of the potential energy) of structures. To that end we run the command
\begin{verbatim}
mlp relax pot.almtp to_relax.cfg relaxed.cfg
\end{verbatim}
The \verb|relax| command can be used both without the control of extrapolation as shown above and with the control of extrapolation. The \verb|--extrapolation_control=true| option turns on calculating the extrapolation grade for each configuration or the extrapolation grade per atom, depending on the active learning mode (configured at the first training). This option also turns on saving either extrapolative periodic configurations or non-periodic parts of the large atomic systems including the extrapolative neighborhoods for which either $\gamma_{\rm save} < \gamma({\rm cfg}) < \gamma_{\rm break}$ or $\gamma_{\rm save} < \gamma(\mathfrak{\bm n}_i^{*}) < \gamma_{\rm break}$. If $\gamma({\rm cfg}) > \gamma_{\rm break}$ (or, $\gamma(\mathfrak{\bm n}_i^{*}) > \gamma_{\rm break}$) then relaxation terminates. The value of the $\gamma_{\rm save}$ threshold can be specified with the option \verb|--threshold_save=<value>|, the value of the $\gamma_{\rm break}$ threshold can be specified with the option \verb|--threshold_break=<value>|, and, finally, the \verb|--save_extrapolative_to=<filename>| option specifies the file for saving the extrapolative (preselected) configurations or, non-periodic configuration parts including the extrapolative neighborhoods. If the \verb|--extrapolation_control=true| option is specified but other related options are missed, then the default values will be used. 

The control of extrapolation allows for preselecting the configurations (neighborhoods) that are the candidates to be added to the training set. For selecting the configurations (neighborhoods) for which the quantum-mechanical calculations will be conducted we use the following command
\begin{widetext}
\begin{verbatim}
mlp select_add pot.almtp train.cfg preselected.cfg selected.cfg
\end{verbatim}
\end{widetext}
The \verb|preselected.cfg| file may contain many configurations (neighborhoods), only the limited number of them will be selected and saved to the \verb|selected.cfg| file. The data for selection (active learning mode and active set) are loaded from the \verb|pot.almtp| file. This command also compares the selected configurations with the ones in the \verb|train.cfg| file, thus we avoid adding the configurations (and conducting expensive quantum-mechanical calculations) similar to what we already have in the training set.

If we use the ``configuration-based'' active learning mode we can directly calculate the periodic configurations from the \verb|selected.cfg| file with any quantum-mechanical package without any additional pre-processing (except for the converting to the format of the concrete quantum-mechanical (DFT) package). However, if we consider the ``neighborhoods-based'' mode then the \verb|selected.cfg| file may contain large non-periodic atomistic structures with different extrapolative neighborhoods. Typically it is not possible to calculate these large atomistic structures with the DFT codes. Therefore we cut the extrapolative neighbohoods from the large configurations by executing
\begin{widetext}
\begin{verbatim}
mlp cut_extrapolative_neighborhood selected.cfg nbh.cfg --cutoff=8 
\end{verbatim}
\end{widetext}
The input file \verb|selected.cfg| contains the column with the grades per atom. This command finds an atom with the largest extrapolation grade and makes the spherical neighborhood \verb|nbh.cfg| including this atom and its neighbors inside the sphere with the given radius \verb|cutoff=8|.

The list and short descriptions of the mentioned commands can be obtained by executing
\begin{verbatim}
mlp list
\end{verbatim}
and 
\begin{verbatim}
mlp help <command>
\end{verbatim}
The commands \verb|cut_extrapolative_neighborhood| and \verb|convert| are serial, the rest of the mentioned commands are capable working with the MPI parallelism.

\subsection{MLIP-3 as a LAMMPS plugin}

MTPs implemented in MLIP-3 can be used as interatomic interaction potentials in LAMMPS. To this end we have to specify the \verb|mlip| ``pair style'' of potentials in the LAMMPS input file:

\begin{verbatim}
pair_style mlip load_from=pot.almtp
pair_coeff * *
\end{verbatim}
where the \verb|load_from| option indicates that the MTP parameters should be loaded from the \verb|pot.almtp| file. As in the \verb|relax| command, we can turn on the extrapolation control while running LAMMPS (e.g., molecular dynamics) and preselect extrapolative configurations (neighborhoods):
\begin{verbatim}
pair_style mlip load_from=pot.almtp 
           extrapolation_control=true \
           threshold_save=2 \ 
           threshold_break=10 \
           save_extrapolative_to=preselected.cfg
pair_coeff * *
\end{verbatim}

\section{Example}
Here we describe how to actively train MTP on atomic neighborhoods starting from one configuration in the training set for simulation of growing a copper crystal on the <111> surface (in what follows we simply refer to this as ``the Cu<111> structure'') by depositing copper atoms on the copper layers (substrate).
The files and scripts needed to run this example available at \verb|https://gitlab.com/ivannovikov/mlip-3-example|.

\subsection{System of Cu<111>}

\begin{figure}
    \centering
    \includegraphics[width=\linewidth]{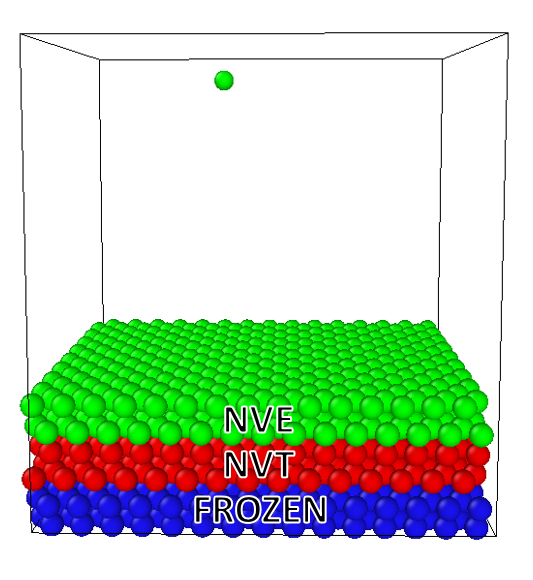}
    \caption{\label{fig:Substrate} Scheme of the substrate and the first deposited atom.}
\end{figure}

Six layers of Cu<111> inside the box of $39 \times 39 \times 40 ~\angstrom$ size (1620 atoms) were taken as a substrate. The distance between the layers was approximately equal to 2.07 $\angstrom$. We took two top layers (and all the deposited atoms) in the NVE ensemble, two middle layers in the NVT ensemble (Nose-Hoover thermostat at temperature $T = 670$ K), and, finally, two bottom layers were fixed along $z$ axis, i.e., the atoms in these layers were allowed to move only along $x$ and $y$ axes. Substrate and the first deposited atom are shown in Fig. \ref{fig:Substrate}.

\subsection{Computational details}

We considered the periodic boundary conditions along $x$ and $y$ axes and non-periodic boundary conditions along $z$ axis. We deposited each atom from a height of 40 $\angstrom$. The MD step was 1 fs. Each atom was deposited in 5000 MD steps ``one-by-one'' with the speed of 1300 meters per second. In each simulation we deposited 1380 atoms.

We actively trained MTPs of 8-th, 12-th, and 16-th levels with 26, 54, and 125 parameters, respectively. We choose $R_{\rm cut}$ = 5.0 $\angstrom$. We fit MTP energies and forces to the ones calculated with DFT, but we did not fit to DFT stresses as we actively selected and added to the training set only the configurations with non-periodic boundary conditions (i.e., spherical neighborhoods). We detected neighborhoods (atoms) as extrapolative starting from the threshold $\gamma_{\rm save} = 2$ (reliable extrapolation, we preselected the neighborhoods with the grades greater than this) until the threshold  $\gamma_{\rm break} = 10$ that corresponds to too high (risky) extrapolation. We break the MD simulation once we detected an atom with this grade or higher, next we selected new neighborhoods among the preselected ones, created the spherical neighborhoods around the atoms with the maximum grades (we took the radius of 8 $\angstrom$), calculated them with DFT, added them to the training set, and re-trained MTP.

We used the VASP package \cite{kresse1993-VASP,kresse1994-VASP,kresse1996A-VASP,kresse1996B-VASP} for DFT calculations. We note that the VASP package operates only with periodic boundary conditions, but we need atoms in a spherical neighborhood to be interacted only inside this neighborhood, not with the ones outside it (i.e., non-periodic boundary conditions). Due to this reason we added vacuum to the selected spherical neighborhoods along each direction. We found that the optimal size of vacuum along each direction is 9 $\angstrom$: starting from this size atoms in different spherical neighborhoods do not interact each other, but interact only inside the same neighborhood. We conducted VASP calculations with the PBE functional \cite{perdew1996-DFT-PBE} and the PAW pseudopotentials \cite{blochl1994-DFT-PAW}. We chose the energy cutoff \verb|ENCUT| = 300 eV and the $1 \times 1 \times 1$ k-mesh.

We note that the details on the convergence of VASP parameters (energy cutoff and k-mesh) and selection of the size of vacuum and the radius of the spherical atomic neighborhood are given in Supplementary Information.

\subsection{Automatic construction of MTP for simulation of copper layers growing}

\begin{figure}
    \centering
    \includegraphics[width=\linewidth]{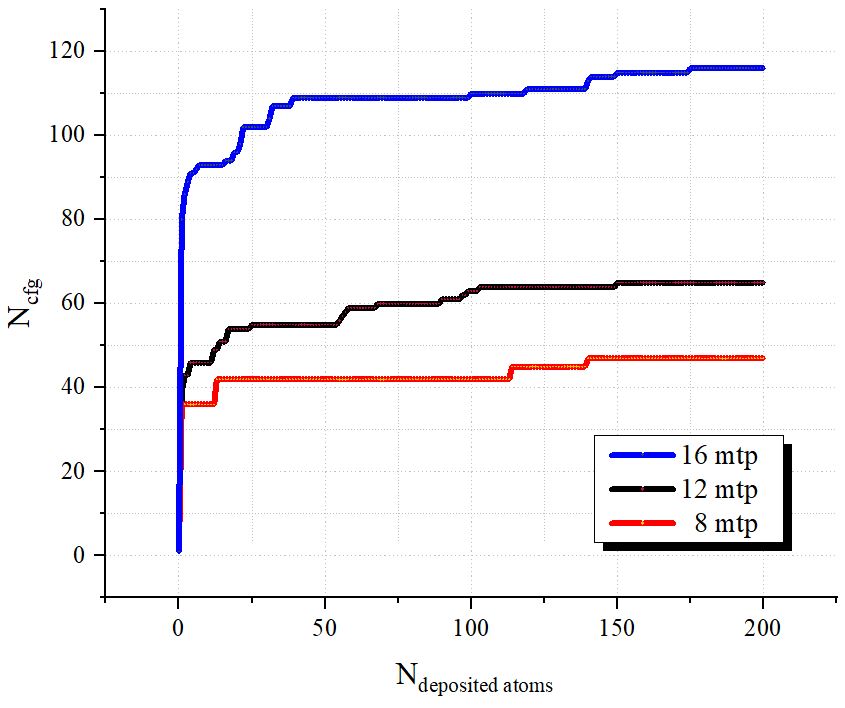}
    \caption{\label{fig:TS_growing} Amount of DFT calculations (or, configurations in the training set) during MD as a function of depositing atoms.}
\end{figure}

We first actively trained MTPs of different levels while depositing atoms on the copper substrate. We demonstrate the process of adding configurations to the training set in Fig. \ref{fig:TS_growing}. For all the potentials we see that for deposition of the first atom we need 30 or more DFT calculations. Most DFT calculations were conducted while depositing the first several atoms. Next, this number continues to grow slowly up to depositing approximately first 200 atoms. We do not need any DFT calculations beyond the deposition of 200 atoms: starting from that point we grow the Cu<111> structure without any DFT calculations. The number of configurations needed to fit MTP increases with the increase in the level of MTP (or, number of parameters). 

\begin{figure}
    \centering
    \includegraphics[width=\linewidth]{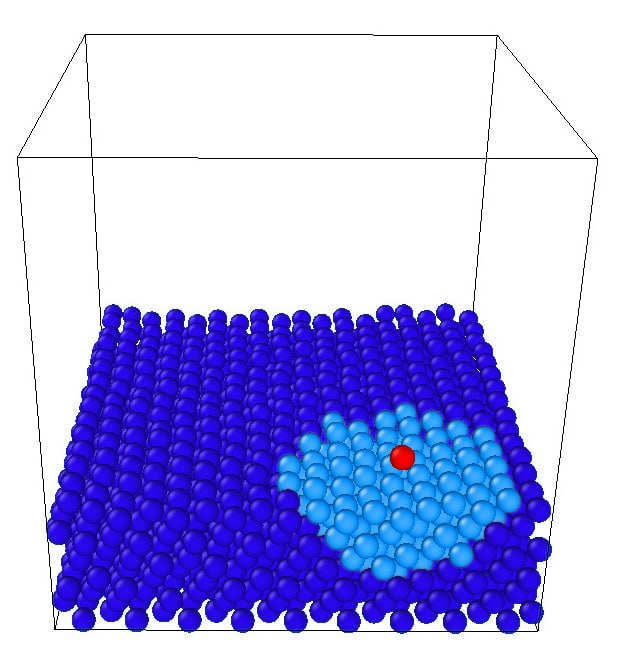}
    \caption{\label{fig:LargeSystem} Large atomic system including candidate extrapolative neighborhoods. The atom with the maximum extrapolation grade and its neighbours are highlighted. These atoms form a spherical neighborhood on which we train the potentials.}
\end{figure}

In Fig. \ref{fig:LargeSystem} we demonstrate an example of a large atomic system containing many candidate extrapolative neighborhoods. In this large system we find an atom with the maximum extrapolation grade and make a spherical neighborhood around it with a given radius. We further add the neighborhoods like this to the training set. 

\begin{figure}[h]
\begin{minipage}[h]{0.32\linewidth}
\center{\includegraphics[width=\linewidth]{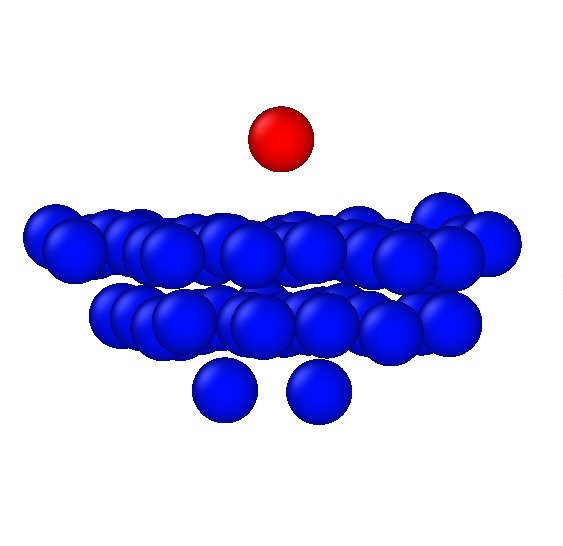} \\ (a)}
\end{minipage}
\hfill
\begin{minipage}[h]{0.32\linewidth}
\center{\includegraphics[width=\linewidth]{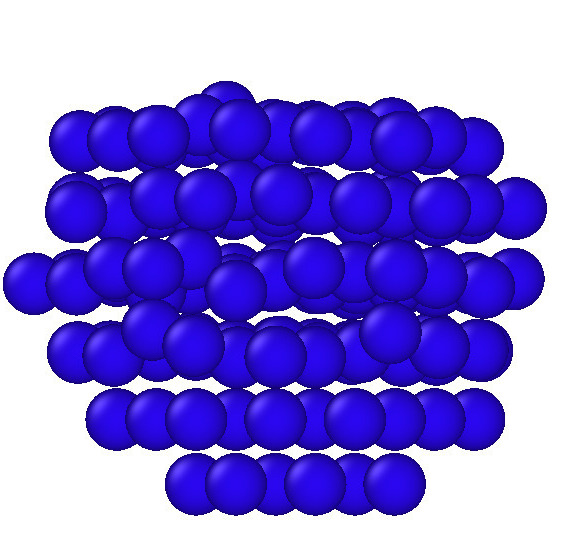} \\ (b)}
\end{minipage}
\hfill
\begin{minipage}[h]{0.32\linewidth}
\center{\includegraphics[width=\linewidth]{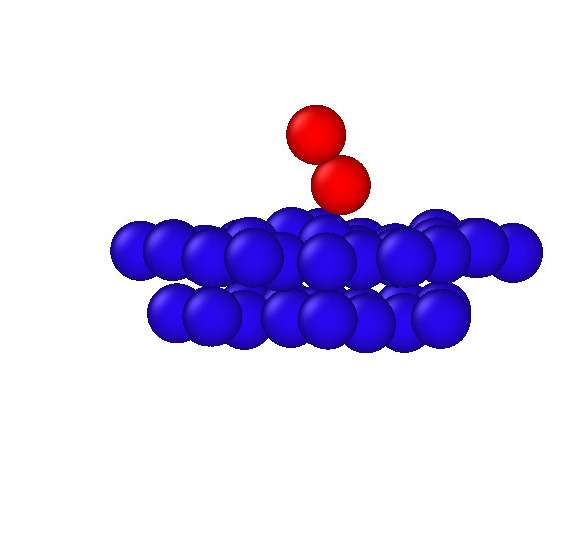} \\ (c)}
\end{minipage}
\caption{\label{fig:SelectedNeighborhoods} Examples of atomic neighborhoods selected during active learning. (a) an atom near the substrate; (b) atoms inside the substrate; (c) several atoms on the substrate. The atoms on the substrate and inside it are of different colors, but all of them are copper.}
\label{ris:image1}
\end{figure}

In Fig. \ref{fig:SelectedNeighborhoods} we show examples of neighborhoods added to the training sets during MD simulations. We first add the neighborhood including one copper atom in vacuum as MTP does not ``recognize'' (or, it extrapolates) such a neighborhood. We next add the neighborhood with the same atom, but near the substrate (a). We also add the neighborhood inside the substrate (b). During the deposition the training set saturates, but some neighborhoods with several atoms on the substrate could be still added (c). 

The resulting training set sizes for the potential of each level are given in Table \ref{tabl:FittingResults}. As was mentioned, training set size increases with the increase in the level of the potential. For estimating the fitting errors and uncertainty of their estimation we consider an ensemble of five potentials for each level. Namely, in addition to the actively trained potentials of different levels we passively trained four MTPs. From Table \ref{tabl:FittingResults} we conclude that the fitting errors decrease with the increase in the level of MTPs. We see that the uncertainty of the fitting errors is rather small. As expected, the highest accuracy was reached with the MTP of the highest considered, 16-th, level. 

\begin{table}
	\begin{center}
		\begin{tabular}{|c|c|c|c|} \hline \hline
			level & Training set size & Energy error, meV/atom & Force error, meV/$\angstrom$ \\ \hline
			8 & 48 & 16.200 $\pm$ 0.003 & 98.878 $\pm$ 0.002 \\ 
			12 & 65 & 10 $\pm$ 1 & 89 $\pm$ 1 \\ 
			16 & 117 & 5 $\pm$ 2 & 68 $\pm$ 1 \\ \hline \hline
		\end{tabular}
		\caption{\label{tabl:FittingResults} Root mean square fitting errors in energy and forces, and training set size actively selected while training the potentials of levels 8, 12, and 16. The 95 \% confidence interval is given for the uncertainty of evaluating the fitting errors based on fitting an ensemble of five potentials. The errors decrease with the increase in the level of potentials, the uncertainty of the errors estimation is small.}
	\end{center}
\end{table}

\begin{figure}
    \centering
    \includegraphics[width=\linewidth]{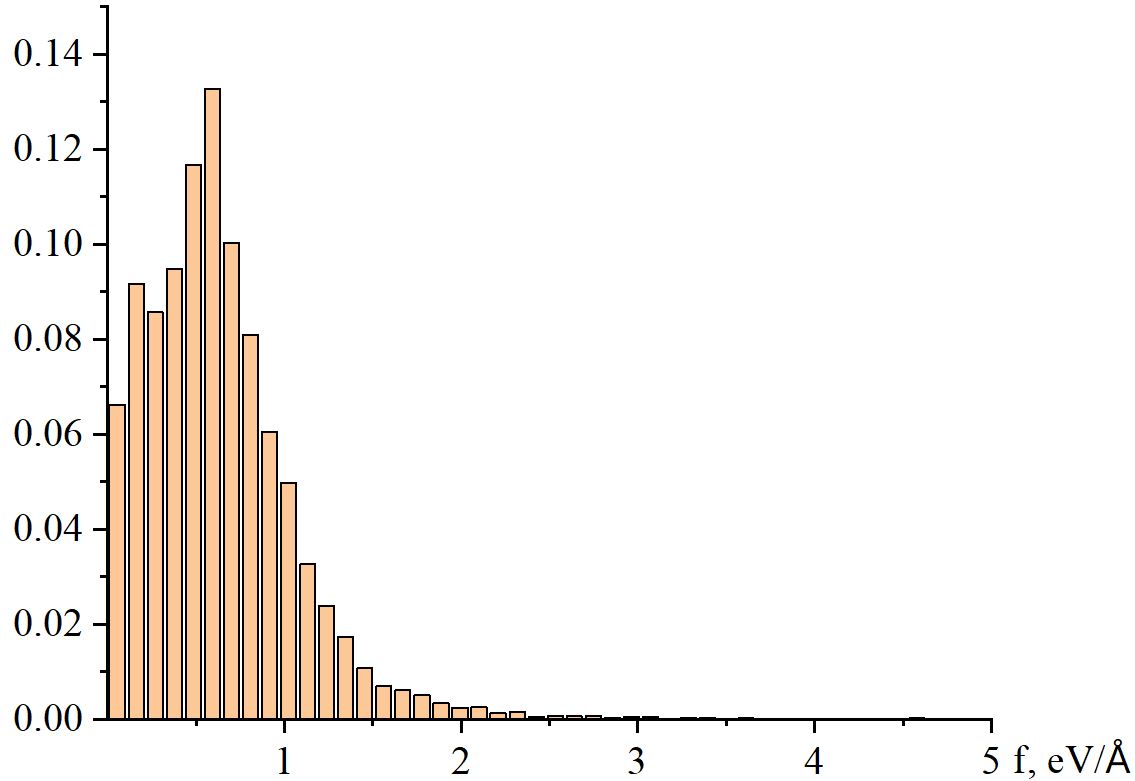}
    \caption{\label{fig:hist} Absolute forces of all the atoms in the training set constructed for the MTP of the 16-th level.}
\end{figure}

We emphasize that we trained MTPs on the forces on all the atoms and the potential energies of the non-periodic neighborhoods. We investigated one of the neighborhoods and found that the force for the central atom in the neighborhood is close to the one for the same atom in the large configuration from which the neighborhood was cut (see Supplementary Information). We also analyzed the forces on all the atoms in the training set automatically created for the MTP of 16-th level and found that almost all absolute values of forces are inside the interval (0,2) eV/$\angstrom$ that is comparable with the forces we have in ordinary periodic configurations (see Fig. \ref{fig:hist} with the histogram).

\begin{figure}
    \centering
    \includegraphics[width=\linewidth]{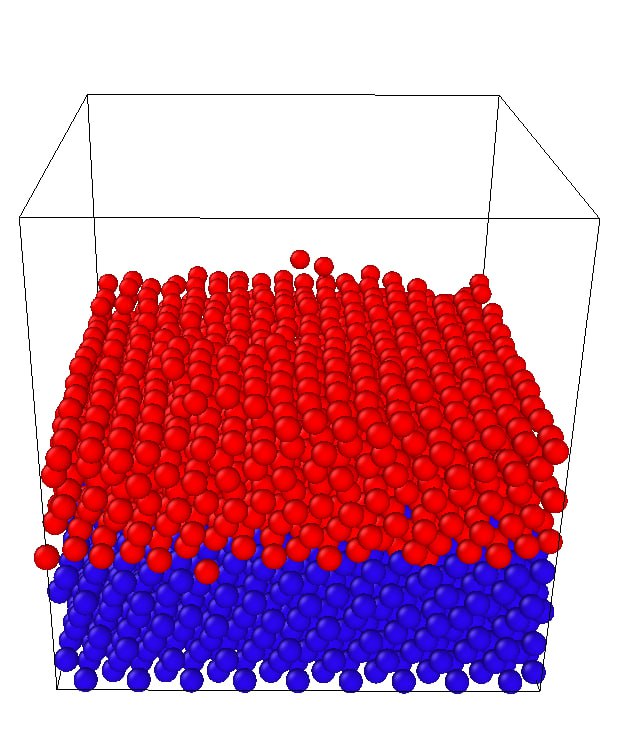}
    \caption{\label{fig:Cu_111_structure} Cu<111> structure grown on Cu<111> substrate with MTP actively trained on atomic neighborhoods.}
\end{figure}

Finally, we demonstrate the Cu<111> structure grown on the Cu<111> substrate with the 16-th level MTP actively trained on atomic neighborhoods in Fig. \ref{fig:Cu_111_structure}. It could be seen that the grown layers have the same structure and ordering as the substrate. We note that we needed only about a hundred of DFT calculations for fitting the MTP which allows us simulating the growth of the Cu<111> structure with the algorithm of active learning on atomic neighborhoods described in this paper.

\section{Conclusions}

An important part of academic research in computational materials science has been the availability of computer codes enabling atomistic modeling, for example the free package LAMMPS \cite{plimpton1995-LAMMPS}, or the commercially available VASP \cite{kresse1993-VASP,kresse1994-VASP,kresse1996A-VASP,kresse1996B-VASP}; most of such codes are developed by academic groups.
The recent developments in the field of machine-learning potentials necessitates for the creation of software packages implementing such potentials.
In this paper we present the MLIP-3 package for constructing and actively training moment tensor potentials.
This package builds on the MLIP-2 package \cite{novikov2020-mlip-2}, but also contains a number of improvements, the most prominent being active learning on atomic neighborhoods of a possibly large atomistic simulation.
This feature enables performing a large simulation, like deposition of atoms on a surface or a nanoindentation simulation \cite{podryabinkin2022-nanohardness}, and training a potential in a completely automatic and autonomous manner, which leads to the possibility of building such an algorithm into complex computational protocols \cite{janssen2019-pyiron,pizzi2016-aiida}.

In this manuscript we described our software package and illustrated its use on a model problem of deposition of Cu atoms on a Cu<111> surface. We hope our work will be useful for simulating atomic deposition processes, especially for the newly developed materials for which a well-tested interatomic potential does not exist and hence the possibility of active learning it while performing the simulation is hence a very attractive option.

\section{Supplementary Material}

Here we provide the details of the convergence of the potential energy with respect to the VASP parameters (energy cutoff and k-mesh), size of vacuum buffer, and radius of the spherical atomic neighborhood.
Additionally, we study the convergence of the force on the central atom in the neighborhood. 

\begin{figure}[h]
\begin{minipage}[h]{0.49\linewidth}
\center{\includegraphics[width=\linewidth]{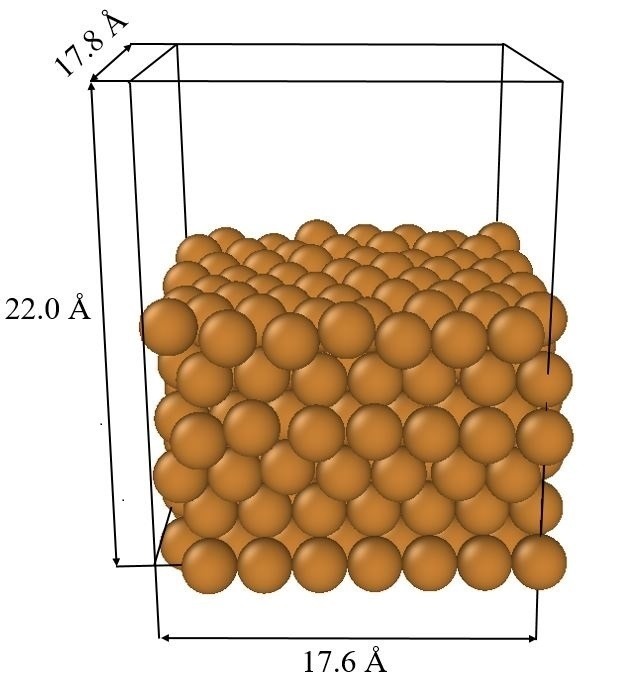} \\ (a)}
\end{minipage}
\hfill
\begin{minipage}[h]{0.49\linewidth}
\center{\includegraphics[width=\linewidth]{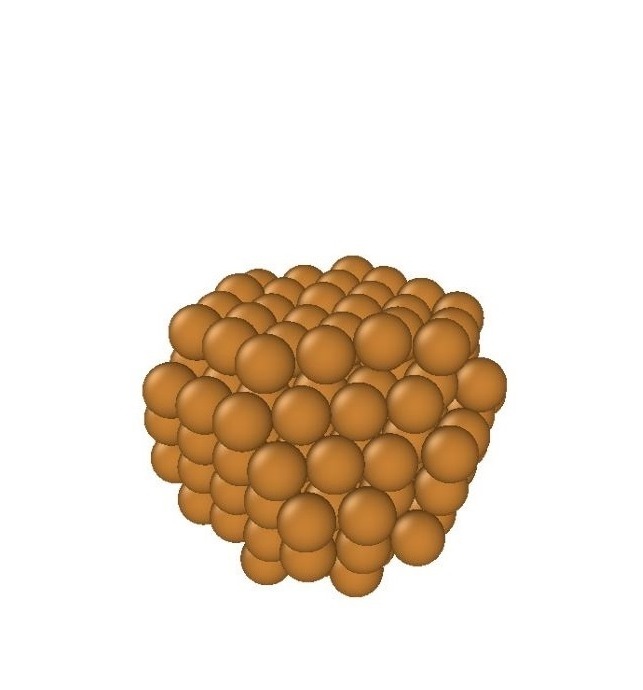} \\ (b)}
\end{minipage}

\caption{\label{fig:cfg_nbh} (a) large configuration of 336 atoms; (b) neighborhood (165 atoms) with the radius of 8 $\angstrom$ that was cut from the large configuration of 336 atoms (a).}
\end{figure}

We started with the VASP parameters for a large configuration with 336 atoms (17.6$\times$17.8$\times$22.0 $\angstrom$) and periodic boundary conditions along the $x$ and $y$ axes (Fig. \ref{fig:cfg_nbh}, (a)). This configuration was obtained during molecular dynamics in the NVT ensemble (Nose-Hoover thermostat at 670 K). 

\begin{table}[h]
	\begin{center}
		\begin{tabular}{|c|c|} \hline \hline
			k-mesh  & Energy, eV \\ \hline
			1$\times$1$\times$1 & $-1182.822$  \\ 
			2$\times$2$\times$1 & $-1181.241$ \\ 
			3$\times$3$\times$1 & $-1181.358$\\ \hline \hline
		\end{tabular}
		\caption{\label{tabl:En_full_conf_kpoints} Convergence of the potential energy with respect to the k-mesh (the dependence of energy of the configuration on k-mesh) for the energy cutoff of 300 eV.}
	\end{center}
\end{table}

\begin{table*}[ht]
	\begin{center}
		\begin{tabular}{|c|c|c|c|c|c|c|} \hline \hline
			\verb|ENCUT|  & 200 & 250 & 300 & 350 & 400 & 450 \\ \hline 
               
    Energy, eV & $-536.152$ & $-1142.060$ &          $-1181.358$ & $-1183.394$ & $-1181.715$ & 
   $-1181.099$\\ 
   \hline \hline
		\end{tabular}
		\caption{\label{tabl:En_full_conf_ENCUT} Convergence of the potential energy of atomic configuration with respect to the energy cutoff for 3$\times$3$\times$1 k-mesh.}
	\end{center}

\end{table*}

\begin{table*}[ht]
	\begin{center}
		\begin{tabular}{|c|c|c|c|c|c|c|c|c|c|c|} \hline \hline
			Size of vacuum, $\angstrom$  & 1 & 2 & 3 & 4 & 5 & 6 & 7 & 8 & 9 & 10 \\ \hline

   Energy, eV & $-421.711$ & $-427.864$ &$-417.697$& $-414.777$ & $-414.163$ &$-414.071$ & $-414.061$ &$-414.044$ & $-414.051$ &$-414.035$  \\ 
 
   \hline \hline
		\end{tabular}
		\caption{\label{tabl:En_box_size} Convergence of the potential energy with respect to the
box size (or, size of vacuum) in which we put the neighborhood (the energy cutoff is 300 eV).}
	\end{center}
\end{table*}

\begin{table}[h]
	\begin{center}
		\begin{tabular}{|c|c|c|c|c|c|c|c|} \hline \hline
			\verb|ENCUT|  & 200 & 250 & 300 & 350 & 400 \\ \hline 
               
    Energy, eV & $-165.849$ & $-399.561$ & $-414.051$ & 
   $-414.762$ & $-414.138$\\ 
   \hline \hline
		\end{tabular}
		\caption{\label{tabl:En_neigh_ENCUT} Convergence of the potential energy of neighborhood with respect to the energy cutoff with the vacuum of 9 $\angstrom$ along each direction.}
	\end{center}

\end{table}

\begin{table}[h]
	\begin{center}
		\begin{tabular}{|c|c|c|c|} \hline \hline
			$r$, $\angstrom$ & $f_x$, eV/$\angstrom$ & $f_y$, eV/$\angstrom$ & $f_z$, eV/$\angstrom$ \\ \hline
 6  & 1.629  & $-0.854$  & $-0.808$ \\
 7  & 1.602  & $-0.909$  & $-0.718$ \\
 8  & 1.649  & $-0.903$  & $-0.795$ \\
 9  & 1.673 &  $-0.891$  & $-0.674$ \\
10  & 1.629 &  $-0.926$  & $-0.769$ \\
11  & 1.597 &  $-0.910$  & $-0.770$ \\
12  & 1.593 &  $-0.923$  & $-0.779$ \\   
			 \hline \hline
		\end{tabular}
		\caption{\label{tabl:forces_neigh} Convergence of the force of the central atom with respect to the size of the carved cluster.}
	\end{center}
\end{table}

To investigate the convergence of VASP calculations with respect to the k-mesh, we fixed the \verb|ENCUT| parameter to 300 eV (which is greater than the energy cutoff in VASP pseudopotential) and performed VASP calculations for different k-meshes (Table \ref{tabl:En_full_conf_kpoints}). The optimal k-mesh is $3 \times 3 \times 1$. Next, we investigated the convergence of VASP calculations with respect to the \verb|ENCUT| parameter (see Table \ref{tabl:En_full_conf_ENCUT}), we took the optimal $3 \times 3 \times 1$ k-mesh.
According to the results of this test we found that the energy cutoff of 300 eV is sufficient for the convergence of VASP calculations for the periodic structure. 

Next, we cut a neighborhood with the radius of 8 $\angstrom$ from this large configuration (Fig. \ref{fig:cfg_nbh}, (b)), took the energy cutoff of 300 eV, 1$\times$1$\times$1 k-mesh, conducted the VASP calculations for various box sizes (Table \ref{tabl:En_box_size}) and found that the box with the size of 17$\times$17$\times$17 $\angstrom$ (i.e., adding ``9 $\angstrom$ of vacuum'' along each direction) is enough for the convergence of the VASP calculations. In other words it guarantees that the atoms interact only inside the atomic neighborhood and we conduct the calculations for the non-periodic cluster. We also provide the table with the convergence of the potential energy with respect to the energy cutoff (we took the vacuum of 9 angstrom) in which we demonstrated that the energy cutoff of 300 eV is enough for the convergence (see Table \ref{tabl:En_neigh_ENCUT}). Thus, we took the energy cutoff of 300 eV, 1$\times$1$\times$1 k-mesh, and added ``9 $\angstrom$ of vacuum'' along each direction for the further VASP calculations of neighborhoods (atomic clusters). 

After investigating the convergence of the potential energy with respect to VASP parameters we investigated the convergence of the force of the central atom with respect to the size of the atomic neighborhood (carved cluster). We took the clusters with the size of 6, 7, 8, 9, 10, 11, and 12 $\angstrom$ and calculated the force (we started from 6 $\angstrom$ to guarantee that the size of the carved cluster is greater than the cutoff radius of machine-learning potential which is equal to 5 $\angstrom$). We found that the force of the central atom does not significantly depend on the size of the atomic cluster (Table \ref{tabl:forces_neigh}). We chose the size of 8 $\angstrom$ for further calculations. We also calculated the force of the same atom in the large configuration with the optimal \verb|ENCUT| and k-mesh and obtained $f_x = 1.628$ eV/$\angstrom$, $f_y = -0.903$ eV/$\angstrom$, and $f_z = -0.773$ eV/$\angstrom$. Thus, we conclude that the force of the central atom in the carved cluster is close to the force of the same atom in the large configuration.

\section{Acknowledgements}

This work was supported by the Russian Science Foundation (grant number 22-73-10206, https://rscf.ru/project/22-73-10206/).

\end{document}